\documentclass[aps,prb,preprint,showpacs]{revtex4}
\usepackage{graphicx}  
\usepackage{amsmath}
\usepackage{amssymb}

\renewcommand{\Im}{\operatorname{Im}}

\begin{document}
\title{Electron transport in the Anderson model}
\author{A.~Alvermann,$^{1,2}$ F. X. Bronold,$^{1}$ and H. Fehske$^{1}$}
\affiliation{$^{1}$Institut f\"ur Physik, Ernst-Moritz-Arndt-Universit\"at
  Greifswald, D-17487 Greifswald, Germany}
\affiliation{$^{2}$Physikalisches Institut, Universit\"at Bayreuth,
 D-95440 Bayreuth, Germany}
\date{\today}
\begin{abstract}
Based on a selfconsistent theory of localization we study the 
electron 
transport properties of a disordered system in the framework of the Anderson
model on a Bethe lattice.
In the calculation of the dc conductivity we separately discuss 
the two contributions
to the current-current correlation function
dominating its behaviour for small and large disorder.
The resulting conductivity abruptly vanishes at a critical disorder strength
marking the localization transition.
\end{abstract}
\maketitle

Disorder strongly affects the motion of an electron,
and even may fully suppress electron transport, as first revealed by 
P.~W.~Anderson~\cite{anderson}. 
The natural quantity to characterize the effect of disorder is the electric
conductivity, whose calculation for a disordered interacting system should be
understood as a primary goal in the field of localization physics.
As a first preparative step, we consider the dc conductivity for a
single (noninteracting) electron in a crystal with compositional disorder.
Our work is directed towards an approach which is applicable for all values of disorder.

The generic model in the field of localization is the Anderson model
\begin{equation}
 H = \sum\limits_i \epsilon_i |i\rangle\langle i| - J \sum\limits_{\langle ij \rangle}
 |i\rangle\langle j | 
\quad . \end{equation} 
Here  $- J \sum\limits_{\langle ij \rangle}|i\rangle\langle j|$ is a
tight-binding term with  hopping matrix element $J$ on a certain lattice
and $\sum\limits_i \epsilon_i |i\rangle\langle i|$ describes random local potentials.
The $\epsilon_i$ are assumed to be identically independently distributed (i.i.d.) variables
with distribution $p(\epsilon_i)=(1/\gamma) \Theta(\gamma/2-|\epsilon_i|)$.
$\gamma$ specifies the strength of the disorder.

For noninteracting particles 
at $T=0$, 
the Fermi function $f(\omega)$ in the Kubo formula 
\begin{equation} \label{eq:kubo}
\sigma_\mathrm{dc} \propto \int d\omega \left(-\frac{df}{d\omega} \right) \chi_{JJ}(\omega)
\end{equation}
for the dc conductivity
becomes a
$\delta$-function, and $\sigma_\mathrm{dc}$ is solely determined by 
the disorder averaged current-current correlation function 
\begin{equation} \label{eq:currcorr}
 \chi_{JJ}(\omega) = \sum\limits_{ij}\sum\limits_{\mathbf{\delta} \mathbf {\delta'}} (-\delta
 \cdot\delta') \langle \Im G_{i,j+\delta'}(\omega) \Im G_{j,i+\delta}(\omega) \rangle_\mathrm{av}
\end{equation}
at the Fermi energy $E_F$
(for interacting particles, this corresponds to a rigid band approximation).
$G_{ij}(\omega)$ denotes the retarted single-particle Green function.
Below, we assume $E_F=0$ (half filled band).

To avoid the need to sum over all lattice sites in
  eq. (\ref{eq:currcorr}),
  Girvin and Jonson \cite{girvin} suggested to split this expression in two
  parts, 
\begin{equation}
 \chi_{JJ}(\omega) = P_1(\omega) \Lambda(\omega) 
 , \end{equation}
with the disorder averaged pair correlation function
\begin{equation}
 P_1 = \langle \Im G_{ii}(\omega) \Im G_{jj}(\omega)-\Im G_{ji}(\omega) \Im G_{ij}(\omega) \rangle_\mathrm{av} 
\end{equation}
for adjacent lattice sites $i,j$,
and a correction $\Lambda(\omega)$ accounting for the long range correlations.
Note that in the regime of localized states $\Lambda(\omega)$ has no definite
value since both $P_1(\omega)$ and $\chi_{JJ}$ vanish.
Close to the localization transition $\chi_{JJ}$ is essentially determined by the pair
correlation function, and $\Lambda(\omega)$ is nearly constant.
For small disorder, on the other hand, $\Lambda(\omega)$ diverges.
To cover the full range of disorder, we must therefore compute both
$P_1(\omega)$ and $\Lambda(\omega)$ to a good approximation.

To calculate the pair correlation function $P_1(\omega)$,
which captures the behaviour near the localization
transition, we employ the self-consistent theory of localization
developed by
Abou-Chacra, Anderson, Thouless (AAT) \cite{abou}.
This theory is based on a renormalized perturbation expansion~\cite{economou}
which sets up a closed set of recursion relations for local Green functions on
the Bethe lattice.
These recursion relations can be interpreted as a stochastic
selfconsistency equation for the Green function $G_{jj}^{(i)}(\omega)$,
which corresponds to the lattice
with site $i$ removed (and appears in the second step of the  renormalized perturbation
expansion).
Solving this stochastic equation  
by a Monte-Carlo procedure 
a sample
for $G_{jj}^{(i)}(\omega)$ is constructed
from which the respective distribution  can be calculated.
On the Bethe lattice the full Green function as well as all nondiagonal Green
functions may be expressed in terms of local Green functions
$G_{jj}^{(i)}(\omega)$ as
\begin{equation}
 G_{ii}(\omega)=\frac{1}{(G^{(j)}_{ii}(\omega))^{-1} - J^2
 G^{(i)}_{jj}(\omega) }
 \quad,\quad\quad
 G_{ij}(\omega)=J G_{ii}(\omega) G^{(i)}_{jj}(\omega)
\quad, \end{equation}
with $j$ nearest neighbour $i$.
Furthermore $P_1(\omega)$ can be calculated from the distribution of $G_{jj}^{(i)}(\omega)$
\begin{equation}
P_1(\omega) = \left\langle \frac{\Im G^{(j)}_{ii}(\omega) \Im G^{(i)}_{jj}(\omega)}{|1-J^2
  G^{(j)}_{ii}(\omega) G^{(i)}_{jj}(\omega)|^2} \right\rangle_\mathrm{av}
\quad.\end{equation}
Note that, since the Bethe lattice has no closed loops, $G^{(i)}_{jj}(\omega)$ and
$G^{(j)}_{ii}(\omega)$ are i.i.d. random variables.
\begin{figure}[htb]
\begin{minipage}[t]{0.45\textwidth}
\includegraphics[width=\textwidth]{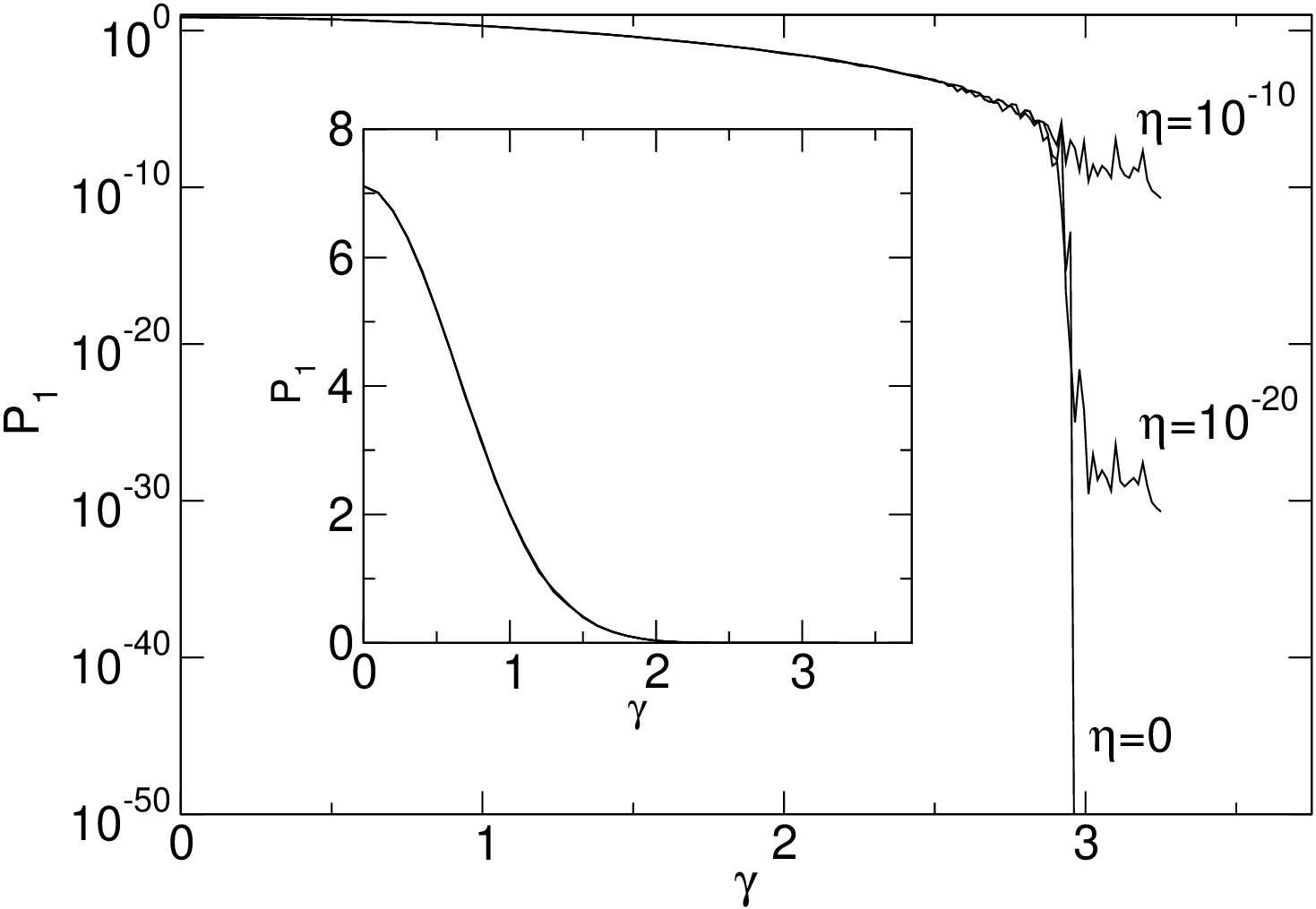}
\caption{Pair correlation $P_1(\omega)$ at the band center ($\omega=0$).
 $\gamma$ is measured in units
of the bandwidth.
$\eta$ denotes the imaginary part of the energy argument in the Green
 functions.}
\label{fig:1}
\end{minipage}
\hfil
\begin{minipage}[t]{0.45\textwidth}
\includegraphics[width=\textwidth]{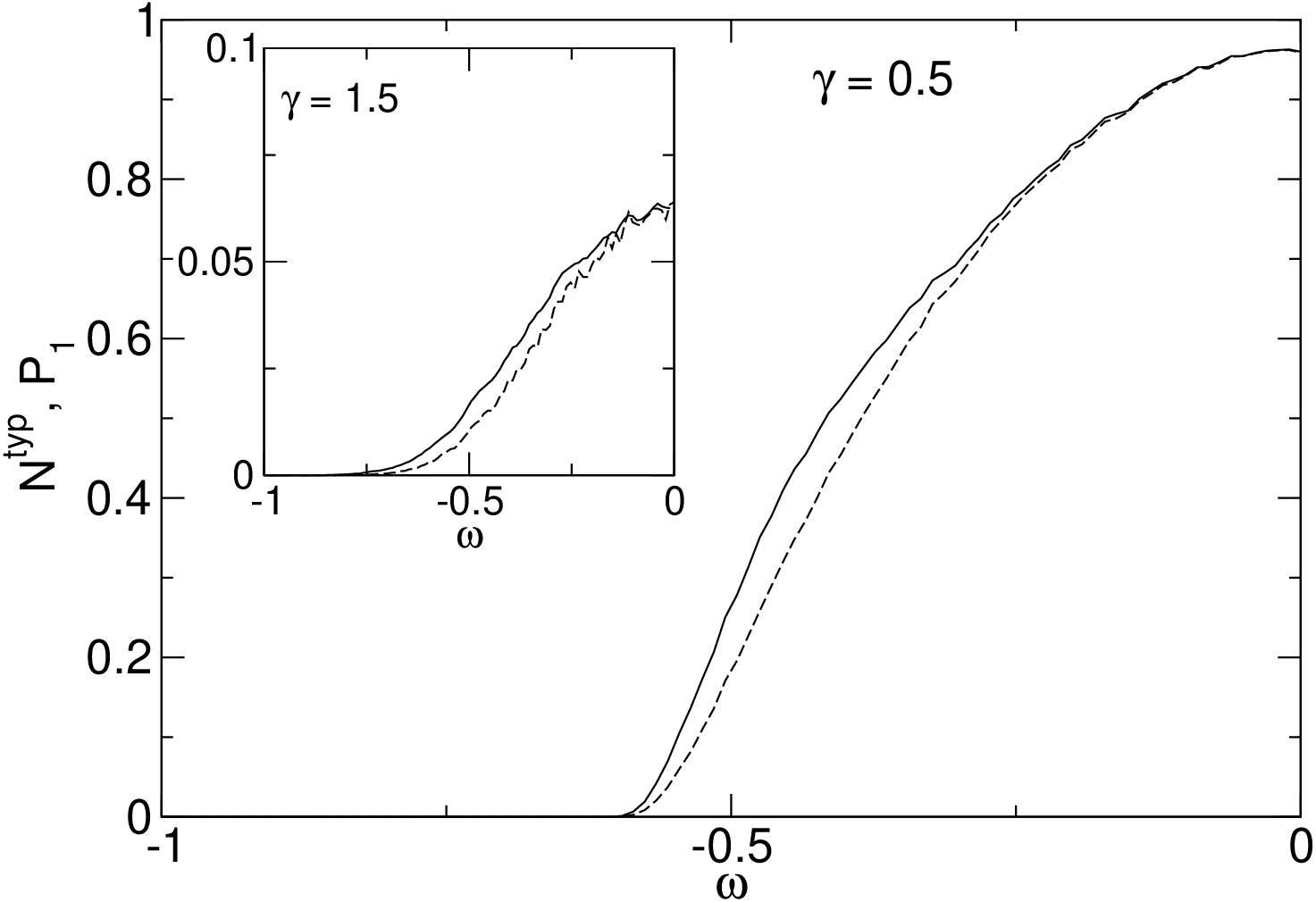}
\caption{Comparison of typical DOS and pair correlations for two
values of $\gamma$.
The solid line shows the pair correlations $P_1(\omega)$, the dashed
 line the typDOS.
 Results are rescaled to their value at $\omega=0$. }
\label{fig:2}
\end{minipage}
\end{figure}

In figure \ref{fig:1} we depict, for the Bethe lattice with connectivity
$K=2$ and for bandwidth $W=1$, the pair correlation function $P_1(\omega)$ at the band
center ($\omega=0)$.
For not too large disorder, $P_1(\omega)$ is finite, but abruptly vanishes for
$\gamma \gtrsim 2.9$,
indicating the transition from extended to localized states.
As figure \ref{fig:1} moreover shows the behaviour of the pair correlation
strongly depends on
the imaginary part $\eta$ in the energy argument $\omega+\mathrm{i}\eta$ of
the Green functions.
To detect the localization transition it is necessary to 
perform the limit $\eta \to 0$ numerically 
(for details on this, see \cite{baf}).

Figure \ref{fig:2} shows a comparison between the pair correlation and
the typical density of states (typDOS)
\begin{equation}
 N^\mathrm{typ}(\omega)=\exp(\langle \ln N^{(i)}_{jj}(\omega) \rangle_\mathrm{av} )
 ,\quad\quad\quad\quad
 \text{ with } \quad N_{jj}(\omega)= - \frac{1}{\pi} G^{(i)}_{jj}(\omega)
,\end{equation}
which has been suggested as an order parameter for localization \cite{dobro}.
Remarkably enough, the pair correlation  follows the
behaviour of the
typDOS over the full range of energy,
 implying a close relation between these two quantities.
In this sense, the typDOS might be itself understood as a kind of 
transport quantity.

\begin{figure}[htb]
\includegraphics[width=0.45\textwidth]{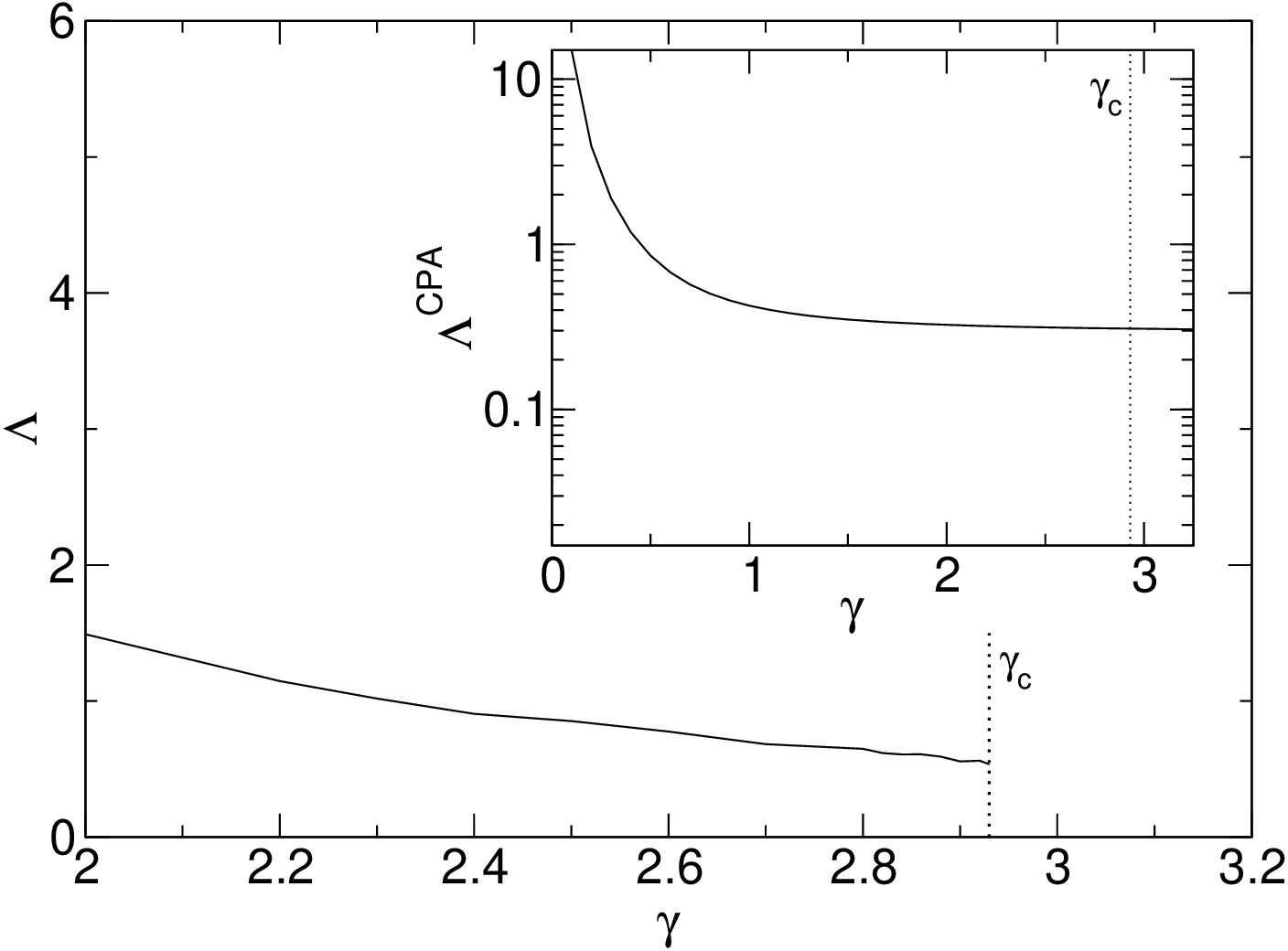}
\caption{$\Lambda(\omega)$-correction to the pair correlation functions
at the band center ($\omega=0$).
The solid line is the AAT-estimate according to eq. (\ref{eq:AATest}), the
dashed line shows the CPA-result.
The vertical dotted lines indicate the position of the localization
transition at  $\gamma_c \approx 2.9$.
 }
\label{fig:3}
\end{figure}

In contrast to $P_1(\omega)$,
the correction $\Lambda(\omega)$  
contains contributions from all terms in eq. (\ref{eq:currcorr}).
Since directions on the Bethe lattice are ill-defined
a direct calculation of these terms within AAT is not possible.
A natural suggestion to overcome this obstacle is to use the coherent potential approximation (CPA).
While for small disorder the CPA surely produces a senseful result
the situation for large disorder is less clear.
To check whether a CPA calculation of $\Lambda(\omega)$ close to the
localization transition is reasonable
we have to estimate the qualitative behaviour of $\Lambda(\omega)$ within AAT.
Running along a single non-retracing path in the Bethe lattice,
 all terms in eq. (\ref{eq:currcorr}) along this path have the form
\begin{equation}
 P_n(\omega) = \langle \Im G_{i,i+n-1}(\omega) \Im G_{i+1,i+n}(\omega)-\Im G_{i,i+n}(\omega) \Im G_{i+1,i+n-1}(\omega) \rangle_\mathrm{av}
\end{equation} 
similarly to the one-dimensional chain
(of course, for $n=1$ we get back the pair correlation function).
On the Bethe lattice $ (K+1)K^{n-1}$ 
paths of length $n$ originate from a single site,
so that the magnitude of $\Lambda(\omega)$ can be estimated as~\cite{girvin}
\begin{equation} \label{eq:AATest}
\Lambda(\omega)\sim\sum\limits_{n=1}^\infty (K+1) K^{n-1} \frac{P_n(\omega)}{P_1(\omega)}
\quad.\end{equation}
Without disorder ($\gamma=0$) each term decays as $P_n(\omega) \sim K^{-n}$,
and the sum diverges.
With disorder, the $P_n(\omega)$ decay faster, resulting in a finite value for $\Lambda(\omega)$.
Figure \ref{fig:3} demonstrates that $\Lambda(\omega)$ is
non-zero and varies only slowly in the vicinity of the localization transition ($\gamma_c \approx 2.9$).
Besides confirming the original expectation, this indicates the possibility 
to employ the CPA for the calculation of $\Lambda(z)$. 
From the CPA current-current correlation function
$\chi_{JJ}^\mathrm{CPA}(\omega)$ \cite{velicky},
 the correction $\Lambda(\omega)$ can be approximated as
$\Lambda^\mathrm{CPA}=\chi_{JJ}^\mathrm{CPA}/P_1^\mathrm{CPA}$
(inset in figure \ref{fig:3}).
That $\Lambda^\mathrm{CPA}(\omega)$ is non-zero even above the localization 
transition makes no problems in our approach:
Since the pair correlation $P_1(\omega)$ is zero, $\chi_{JJ}(\omega)$
is as well, indicating localization of states.

\clearpage

Now we are in the position to put pieces together:
If we calculate the pair correlation function $P_1(\omega)$ within AAT, and
the correction $\Lambda(\omega)$ within CPA, 
the current-current correlation function $\chi_{JJ}(\omega)$ 
is expected to be obtained in good approximation.
Although it is not necessary to split  the
current-current correlation function in the proposed way,
it seems quite reasonable.
First, we have identified the two contributions to the current-current
correlation function which dominate its behaviour in the regimes of
small respective large disorder. 
Second, since both contributions show different behaviour
we can apply different methods for their calculation.

The pair correlation function $P_1(\omega)$ rules the behaviour of
$\chi_{JJ}(\omega)$ for large disorder.
While it is finite for all values of disorder, it is critical at the
localization transition. 
Indeed the full effect of localization is contained 
 in $P_1(\omega)$ but not in $\Lambda(z)$.
Mean field treatments like the CPA, merely focusing on average values,
are not sufficient for $P_1(\omega)$.
The apparent relationship between the typical DOS and the pair correlation function makes this point 
especially clear:
The pair correlation function does not agree
with the averaged DOS (``arithmetic mean''), but with the typDOS 
(``geometric mean''). 
This failure of mean field treatments indicates the subtlety of the
problem at hand.
In constrast the selfconsistent theory of localization by Abou-Chacra
et.~al.~\cite{abou} can be conveniently applied, and
allows for a not too complicated calculation of the pair correlations.
The correction $\Lambda(\omega)$ is in some sense opposite to 
the pair correlation function.
While it is noncritical at the localization transition, it diverges for zero
disorder. 
If one is only interested in the behaviour close to the localization transition
one might completely forget about $\Lambda(\omega)$.
But to correctly describe the regime of small disorder 
$\Lambda(\omega)$
has to be taken into account.
Nevertheless, since $\Lambda(\omega)$ shows no critical behaviour,
there is no need to go beyond a mean field treatment, e.g. provided by the CPA.
As additional benefit, this avoids all possible problems arising from
the lack of well defined directions on the Bethe lattice.

In conclusion, the proposed splitting of the current-current correlation function
seems to be a successful first approximation.
Since the pair correlation $P_1(\omega)$ is a transport quantity 
it is a natural localization criterion, whose
calculation allows 
for a precise determination of mobility edges and the critical disorder
for the localization transition.
The relation between $P_1(\omega)$  and the typDOS shows that 
the latter one can indeed be used as a localization criterion as
in~\cite{baf,statDMFA,BF02}. 
The $\Lambda(\omega)$-correction, on the other hand, is of no importance
to detect the localization transition, but accounts for the divergence of the
current-current correlation function as $\gamma\to 0$.
The final result for $\chi_{JJ}(\omega)$
correctly interpolates between the two limiting cases $\gamma \to 0$ and
$\gamma \to \gamma_c$, where it matches the exact result.

Note that the AAT and the CPA can be combined to some extent with a treatment
of interaction processes (statDMFA, DMFA/DCPA).
Adopting the ideas presented here to these extensions 
might hint at some possible direction for future attempts to 
deal with electron transport in a disordered interacting system.
Especially the selfconsistent theory of localization~\cite{abou}, which can be straightforwardly combined with a
local treatment of interaction (statDMFA \cite{baf,statDMFA,BF02}), is a promising
candidate for these studies.
Work on this subject is in progress.

One of the authors (A.A) acknowledges support from the European Graduate
 School, Bayreuth.

\end{document}